# Inelastic p$^9$Be scattering and halo-structure of excited states of $^9$Be

E.T. Ibraeva[1,*], M.A. Zhusupov[2], A.V. Dzhazairov-Kakhramanov,[1,3,†] P.M. Krassovitskiy[1]

[1] *Institute of nuclear physics RK, 050032, str. Ibragimova 1, Almaty, Kazakhstan*
[2] *Al-Farabi Kazakh National University, 050040, av. Al-Farabi 71, Almaty, Kazakhstan*
[3] *V. G. Fessenkov Astrophysical Institute "NCSRT" NSA RK, 050020, Observatory 23, Kamenskoe plato, Almaty, Kazakhstan*
[*] ibraeva.elena@gmail.com
[†] albert-j@yandex.ru

**Abstract**

The calculation of differential cross section of inelastic p$^9$Be scattering (to the levels $J^\pi = 1/2^+, 3/2^+$) was made in the framework of the Glauber diffraction theory. We have used the wave function of $^9$Be in the ground and excited states in the three-body $2\alpha$n model. Expansion in series by gaussoids of the wave function of $^9$Be and presentation of the Glauber's operator $\Omega$ in the form, conjugated with three-body wave function. It allows us to analytically calculate the matrix elements of inelastic scattering, taking into account all of the multiplicities of scattering and rescattering on clusters and nucleon, that are the components of $^9$Be. The drawn up profiles of excited state functions allow us to make conclusion on their extended neutron distribution. The differential cross section with the wave function in model 1 (with the $\alpha\alpha$-Ali-Bodmer potential) is in a good agreement with available experimental data at $E = 180$ MeV.

## 1. Introduction

New stimulus for the research on weakly bound nuclei was the discovery of the exotic structure (halo and skin) of the series of unstable neutron- and proton-rich isotopes. The measurement of elastic scattering and reactions (breakup of light nucleus in the field of heavy one, stripping/transfer of weakly bound nucleon or cluster) in experiments involving such nuclei, indicate that these processes offer broad opportunities to understand the underlying cluster structure effects.

$^9$Be is a stable nucleus, strongly deformed (quadrupole moment $Q = 52.88(38)$ mb) [1] and the weakly bound in $^9$Be $\to$ $^8$Be + n ($\varepsilon = 1.67$ MeV) [1] and $^9$Be $\to \alpha + \alpha + $ n ($\varepsilon = 1.57$ MeV) [1] channels, which is a direct indication of its three-body $\alpha+\alpha+$n structure. The wave functions (WFs) taking into account this fact, calculated in $2\alpha$n model [2−5], provide a good description of its static observables ($\langle r^2 \rangle^{1/2}$, $Q$, $\mu$), electromagnetic form factors and characteristics of the elastic $\pi$-, K- and p-$^9$Be scattering [5−8].

This nucleus − is a Borromian nucleus, because within the three-body $2\alpha$n picture of $^9$Be there are no two components that could form a bound system, as in the case of $^6$He and $^{11}$Li. Although, the value of the root mean square (rms) radius ($\langle r^2 \rangle^{1/2} = 2.45(1)$ fm) [1] does not indicate the halo structure in this nucleus in its ground state, but in excited states $J^\pi = 1/2^+, 3/2^+$ the rms radii are greater: 2.83 fm and 2.98 fm, respectively. It seems that being excited, the nucleus increases in size and changes his structure to halo. As it is shown below (under review of

WF configurations) and as noted in [2, 3], "the state $J^\pi = 1/2^+$ is very diffuse near-threshold state. The correct interpretation of the three-body asymptotic behavior of the $1/2^+$ WF is important". Increase of the radius (by 1.2 fm) in this state is derived in the experiment on the inelastic $\alpha^9$Be scattering. This experiment was conducted at University of Tsukuba's tandem accelerator at $E_\alpha = 30$ MeV [9, 10].

Monte Carlo's calculation of ground and low-lying excited states of nuclei with $A = 9$ their static characteristics and densities were carried out in [11]. The realistic two- (AV18) and three-body (UIX, Il1-Il5) Hamiltonians were used. It was found that Hamiltonian with Illinois three-nucleon potentials reproduces 10 states of $^9$Li, $^9$Be, $^{10}$Be and $^{10}$B with the rms deviation less than 900 keV. A good agreement was also obtained between the calculated static characteristics (neutron and proton rms radii, magnetic and quadrupole moments) with the experimental ones.

Recent data on the study of certain characteristics (the binding energy, the charge radius, the parameter of quadrupole deformation) are presented in [12]. The Glauber model was used to describe the dynamics of the reaction with the densities, it was obtained from two formalisms: microscopic nonrelativistic Hartree-Fock and relativistic mean field. A good agreement was achieved with the experimental values for both formalisms.

The cluster structure of $^9$Be is interesting not only for studying the dynamics of reactions with weakly bound nuclei. It is important for nuclear astrophysics, because it is the most effective way to bridge the energy "gap" in nucleosynthesis at A = 5 and A = 8 is the $^9$Be($\alpha$, n)$^{12}$C reaction, which follows from the $(\alpha + \alpha) + n \rightarrow {}^8$Be $+ n \rightarrow {}^9$Be reaction. For calculating the reaction rates it is necessary to know the cluster structure of $^9$Be. Therefore, although the three-body $\alpha + \alpha + n$ picture of $^9$Be is most probable, the effective two-body $\alpha + {}^5$He [$\alpha + (\alpha + n)$] or $n + {}^8$Be [$n + (\alpha + \alpha)$] cluster configurations are also considered in the process of $^9$Be breakup in the field of heavy nucleus or nucleon transfer in $^9$Be($^3$He, $\alpha$)$^8$Be [4, 13]. Thus, the $^9$Be cluster structure in channels $^5$He + $^4$He and $^8$Be + n is discussed for the description of the available data on the elastic $^9$Be scattering on $^{208}$Pb [13−15], $^{28}$Si, $^{64}$Zn, $^{144}$Sm [16], $^{186}$W [17] targets at low energies in the region of the Coulomb barrier. The effects of coupling channel are also considered during nucleon breakup and transfer. The results of these calculations show that the coupling effects during breakup are important for the $^5$He + $^4$He cluster model at the energies above the Coulomb barrier. At the same time for the lower (sub-Coulomb barrier energies) it is necessary to consider the $^8$Be + n configuration. It is shown in [14] that only the $^8$Be + n cluster structure of $^9$Be provides satisfactory explanation of data of the $^9$Be+$^{208}$Pb scattering. In the recent high precision experiment [15] the measuring of the cross section for elastic the $^9$Be+$^{208}$Pb scattering at the sub-barrier energies revealed, that the observed deviation from the Rutherford scattering points to dominating the $^8$Be + n cluster structure of $^9$Be, whereas the $\alpha + {}^5$He structure was not confirmed.

The different theoretical methods: the laser spectroscopy [18], the distorted waves method [12−15, 19−21], the coupled channel method [19], and the Glauber's multiple diffraction scattering [5−9, 22, 23] were used for description of the properties of $^9$Be and scattering characteristics of protons. Thus, the $^9$Be charge radius is equal to 2.519 (12) fm [18] and measured by the method of laser spectroscopy, which is the most accurate method at present moment. Elastic and inelastic (to the level $J^\pi = 5/2^-$, $E^* = 2.44$ MeV) scattering of polarized protons at the energy of 220 MeV was previously measured in [19]. Calculation of the differential cross section, the analyzing power, and the depolarization was made in the optical model, in the DWBA and in the coupled channel method using the spherical Woods-Saxon potential. The results show, that the simple optical model and the DWBA provide worse description of the cross section and the polarization characteristics in first approximation than the coupled channel method.

Differential cross sections and analyzing powers of the p$^9$Be scattering and the $^9$Be(p, n)$^9$B charge-exchange reaction at $E = 180$ MeV for the ground and the excited states of $^9$Be are calculated in the DWBA using the effective interaction. It depends on a density and based on the Paris potential [20]. Comparison with the experimental data shows that taking into account the quadrupole deformation of $^9$Be the differential cross sections are well reproduced either for primary or for almost all excited states in a wide range of momentum transfers of $q = 0$–$3$ fm$^{-1}$, while the analyzing power – is slightly worse.

Systematic study of the elastic scattering of protons on some nuclei from $^6$Li to $^{208}$Pb (including $^9$Be) in the full microscopic model with nonlocal optical potential was made in [21] at $E = 200$ MeV. A good agreement with the available experimental data was obtained for the differential cross sections and the analyzing power for the p$^9$Be scattering.

The study of inelastic scattering of α-particles on $^9$Be and one-body transfer reactions $^9$Be(α, $^3$He)$^{10}$Be and $^9$Be(α, $^3$H)$^{10}$B has been recently performed in Finland (Cyclotron Facility of the Accelerator Laboratory Jyvaskayla University) at $E_α = 63$ MeV. The measured differential cross sections for ground and some low-lying (5/2$^-$, 7/2$^-$, 9/2$^-$) states were analyzed within the framework of the optical model, the coupled channel method and the DWBA [24].

The pioneer works of Alkhasov [22, 23] containing the measured differential cross sections of protons scattering on nuclei $^9$Be, $^{11}$B, $^{12,13}$C, $^{14}$N, $^{16}$O at $E = 1$ GeV. This was a stimulus for further study of elastic and inelastic scattering of protons. The analysis of experimental data was made in the Glauber theory of multiple diffraction scattering; meanwhile the parameters of spherical and non-spherical component density were extracted for the abovementioned nuclei.

In our previous works [6−8], elastic and inelastic differential cross section (to the level $J^π = 1/2^+$) within the Glauber theory are calculated at $E = 220$ MeV and compared with the experimental data [19], where this level with $E^* = 1.68$ MeV was not allowed. The reason was that the accuracy of the recording equipment does not exceed 2 MeV, therefore comparison with the experiment was made only for scattering for the ground state of $^9$Be.

In order to find out how the diffuse structure of the $^9$Be excited state effect the inelastic scattering of protons, we have calculated the differential cross section at $E = 180$ MeV and compared with the experimental data [20], where $J^π = 1/2^+$ level is allowed.

## 2. Wave function of $^9$Be in 2αn model

Because the nucleus $^9$Be is strongly deformed, its structure and the energy spectrum cannot be identified quite accurate in the shell model [25]. The fact of distortion of the shell structure is indicated by such facts, as pygmy resonance in the cross section of photo-absorption on $^9$Be at $E^* = 4$ MeV and abnormally small distance between the levels of opposite parity.

Sequential calculation of the WF of $^9$Be was done in [2, 3] in the framework of three-body 2αn model (see Fig. 1) with the paired αn and αα interactions, involving the states forbidden by the Pauli principle, reflecting the composite nature of α particles. The role of the Pauli principle in the description of nuclei structure is, that it does not allow a strong overlap of valent nucleon with α particles, as well as the overlap of two α particles with each other and thus greatly reduces the influence of the internal area of αn and αα interactions.

The wave function of $^9$Be with total angular momentum $J$ and its projection $M_J$ can be written as follows [2]:

$$\Psi_{i,f}^{JM_J} = \varphi_{Jα=Tα=0}(\mathbf{r}_1,...\mathbf{r}_4)\varphi_{Jα=Tα=0}(\mathbf{r}_5,...\mathbf{r}_8)\Psi^{JM_J}(\mathbf{r},\mathbf{R}), \qquad (1)$$

where $\varphi_{Jα=Tα=0}$ is the WF of the α particle, depending on internal coordinates of the four-nucleons system.

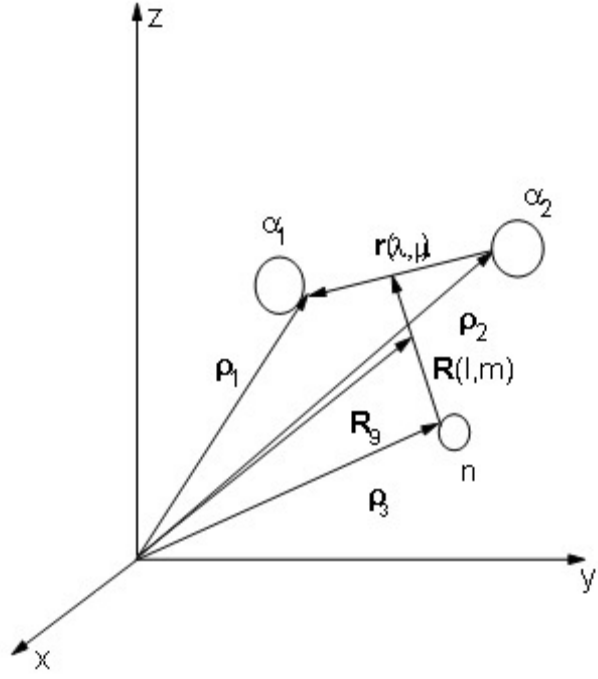

Fig. 1. Scheme of $^9$Be in the 2αn model. $\rho_1$, $\rho_2$, $\rho_3$ – single-body coordinates of clusters and the nucleon, **R**, **r** – Jacobi coordinates, $l$, $\lambda$ – momentums conjugated with them.

In the ground state ($J^\pi = 3/2^-$) the contribution is provided by three components with approximately equal weights:

$$\Psi^{JM_J}(\mathbf{r},\mathbf{R}) = \Psi^{\lambda lL}_{011}(\mathbf{r},\mathbf{R}) + \Psi^{\lambda lL}_{211}(\mathbf{r},\mathbf{R}) + \Psi^{\lambda lL}_{212}(\mathbf{r},\mathbf{R}). \tag{2}$$

where **r** and **R** are Jacobi coordinates, $\lambda$ and $l$ are the conjugated orbital momenta, $L$ is the total orbital angular momentum of the nucleus, $\boldsymbol{\lambda} + \mathbf{l} = \mathbf{L}$. Wave function components (2) are written as follows [2]:

$$\Psi^{\lambda lL}_{011}(\mathbf{r},\mathbf{R}) = \frac{1}{\sqrt{4\pi}} \sum_{M_L M_S} \left\langle 1 M_L \frac{1}{2} M_S \bigg| \frac{3}{2} M_J \right\rangle \delta_{m M_L} \cdot R \cdot Y_{1m}(\Omega_R) \chi_{\frac{1}{2} M_S} \sum_{ij} C^{01}_{ij} \exp(-\alpha_i r^2 - \beta_j R^2), \tag{3}$$

$$\Psi^{\lambda lL}_{211}(\mathbf{r},\mathbf{R}) = \sum_{M_L M_S} \left\langle 1 M_L \frac{1}{2} M_S \bigg| \frac{3}{2} M_J \right\rangle \langle 2\mu 1m | 1 M_L \rangle r^2 Y_{2\mu}(\Omega_r) R Y_{1m}(\Omega_R) \chi_{\frac{1}{2} M_S} \times \\ \sum_{\nu\varepsilon} C^{21}_{\nu\varepsilon} \exp(-\alpha_\nu r^2 - \beta_\varepsilon R^2), \tag{4}$$

$$\Psi^{\lambda lL}_{212}(\mathbf{r},\mathbf{R}) = \sum_{M_L M_S} \left\langle 2 M_L \frac{1}{2} M_S \bigg| \frac{3}{2} M_J \right\rangle \langle 2\mu 1m | 2 M_L \rangle r^2 Y_{2\mu}(\Omega_r) R Y_{1m}(\Omega_R) \chi_{\frac{1}{2} M_S} \times \\ \times \sum_{\gamma\zeta} C^{21}_{\gamma\zeta} \exp(-\alpha_\gamma r^2 - \beta_\zeta R^2), \tag{5}$$

where $\left\langle L M_L \frac{1}{2} M_S \big| \frac{3}{2} M_J \right\rangle$, $\langle 2\mu 1m | 2 M_L \rangle$ are Clebsch-Gordan coefficients, which determine the coupling scheme of momenta, $Y_{\lambda\mu}(\Omega_r)$, $Y_{lm}(\Omega_R)$ are spherical functions, $\chi_{SM_S}$ is the spin function, $C^{\lambda l}_{ij}$, $\alpha_i$, $\beta_j$ are linear and nonlinear variational parameters. Relative weights of three WF configurations and some static characteristics of $^9$Be are shown in the following table.

Table. The considered configurations, their relative weights in the $^9$Be WF and the static characteristics calculated in [2−4]: the rms charge radius $r_{ch}$, quadrupole $Q$ and magnetic $\mu$ momenta.

| colspan="5" | Ground state of $^9$Be, $J^\pi = 3/2^-$ |||||
|---|---|---|---|---|
| $\lambda$ | $l$ | $L$ | Model 1 (AB) | Model 2 (BFW) |
| 0 | 1 | 1 | 0.404 | 0.439 |
| 2 | 1 | 1 | 0.338 | 0.355 |
| 2 | 1 | 2 | 0.235 | 0.196 |
| colspan="3" | $r_{ch}^*$, fm ||| 2.52 | 2.34 |
| colspan="3" | $Q^{**}$, mb ||| 50.0 | 38.0 |
| colspan="3" | $\mu^{***}$, $\mu_0$ ||| -0.854 | -0.947 |
| colspan="5" | $^*r_{ch.\ exp.} = 2.519(12)$ fm [1] |||||
| colspan="5" | $^{**}Q_{exp.} = 53\pm 3$ mb [1] |||||
| colspan="5" | $^{***}\mu_{exp.} = -1.1778(9)\mu_0$ [1] |||||
| colspan="5" | Excited state of $^9$Be, $J^\pi = 1/2^+$ |||||
| $\lambda$ | $l$ | $L$ | Model 1 (AB) | Model 2 (BFW) |
|  | 0 | 0 | 0.911 | 0.993 |
| colspan="5" | Excited state of $^9$Be, $J^\pi = 3/2^+$ |||||
| $\lambda$ | $l$ | $L$ | Model 1 (AB) | Model 2 (BFW) |
| 0 | 2 | 2 | 0.995 | 0.996 |

The low-lying excited states of $^9$Be ($J^\pi = 1/2^+, 3/2^+$) contain one dominant component. Let us give the WF form, used in the calculation. The state $J^\pi = 1/2^+$ with $E^* = 1.68$ MeV and with relative weight 99.7% [2]:

$$\Psi_{000}^{\lambda lL}(r,R) = \frac{1}{\sqrt{4\pi}} \chi_{\frac{1}{2}M_S} \sum_{ij} C_{ij}^{00} \exp(-\alpha_i r^2 - \beta_j R^2) . \qquad (6)$$

The state $J^\pi = 3/2^+$ with $E^* = 4.704$ MeV [1] with weight 97.9% [2]:

$$\Psi_{022}^{\lambda lL}(\mathbf{r},\mathbf{R}) = \sum_{M_L M_S} \left\langle 2M_L \frac{1}{2} M_S \bigg| \frac{3}{2} M_J \right\rangle \langle 002m | 2M_L \rangle Y_{00}(\Omega_r) R^2 Y_{2m}(\Omega_R) \chi_{\frac{1}{2}M_S} \times$$
$$\times \sum_{\gamma\zeta} C_{ij}^{02} \exp(-\alpha_i r^2 - \beta_j R^2).$$

The calculation of WF in [2, 3] was made using the variational stochastic method with three pair interactions $V_{\alpha\alpha}, V_{\alpha_1 n}, V_{\alpha_2 n}$. The following was used as the pair interactions.

Model 1: $V_{\alpha\alpha}$ is the Ali-Bodmer (AB) potential [26], which is shallow with the repulsive core at small distances, and does not contain forbidden states. $V_{\alpha n}$ – the potential with exchange Majorana component, which leads to an even-odd splitting of the phase shifts and better (than the previously used Suck-Biedenharn-Breit (SBB) potential) reproduces the phases with $l = 0, 1$ and 2.

Model 2: $V_{\alpha\alpha}$ – the Buck-Friedrich-Wheatley (BFW) potential [27], the deep attractive potential with the Pauli forbidden states, describing the scattering phase with $\lambda = 0, 2, 4$ и 6; $V_{\alpha n}$ – the same as in model 1.

Studying the role of αα interaction for calculation of the WF of $^9$Be [2, 3], the composite structure of α particles was taken into account through the WF of relative αα movement by the two fold manner. In model 1, the accounting of the Pauli principle is associated with the use of $\lambda$ dependent AB potential, containing repulsion at small distances. In model 2, the deep attractive potential with the BFW states, forbidden by the Pauli principle was used. In this potential the WF of the 4$S$-shell type function, that containing two nodes corresponds to the state with orbital momentum $\lambda = 0$. The functions with 0$S$ and 2$S$ correspond to the forbidden states and in three-body calculations they are excluded by using a special projection procedure [2, 3].

What is the difference between WFs calculated in different potentials? The WF of the ground state in model 1 is close to zero because of the effect of repulsive core inside the nucleus ("disappears"), and reaches the maximum value at the periphery, at $r > 3-4$ fm (see Fig. 9 below). The values for this WF, which operators grow with distance (rms charge radius and quadrupole momentum), have higher values than in model 2, where the WF inside the nucleus is not dying but it oscillates. On the contrary, the values determined by the entire nuclear volume (magnetic momentum, neutron spectroscopic factors) have larger values in model 2, where the WF is more drawn into the nucleus, and has two maxima and the node in the inner region. The energy spectrum in model 1 is transmitted better: the ground state is rebounded only by 0.2 MeV (in model 2 – by 1.5 MeV), level 1/2$^+$ is overrated by 0.5 MeV (in model 2 – by 1 MeV). Longitudinal electromagnetic form factors in both models are approximately the same [2–4].

Thus, the static characteristics and the energies of the ground and excited states of $^9$Be are better described by model 1, where the AB αα-potential, containing repulsion at small distances, is used for the WF calculation.

Therefore the authors [2, 3] conclude: "the true nonlocal αα interaction is probably close to the shallow $\lambda$ dependent potential (AB), but not to the deep $\lambda$ dependant potential (BFW). In particular, the amplitude of the WF of αα relative movement must be suppressed in the region of clusters overlap (as it takes place for the AB potential). Therefore the correct picture of the αα interaction should be similar to that given by the Resonating Group Method (RGM) and by the Generator Coordinate Method (GCM) with strongly suppressed oscillations of the relative movement WF in the internal region due to the Pauli principle".

Lets consider the geometric structure of model WFs, which makes possible to visualize the relative location of clusters and to understand the demonstration of their properties in the scattering process.

Figs. 2−9 show the behavior of the function profiles $W(r,R) = \sum_{\lambda,l,L} |\Psi^{\lambda l L}|^2 r^2 R^2$ of the excited $J^\pi = 1/2^+$, $3/2^+$ and ground $J^\pi = 3/2^-$ states of $^9$Be.

Lets consider the profiles of functions of the first excited state $J^\pi = 1/2^+$. The three-dimensional profiles of the WF in the potentials AB (Fig. 2) and BFW (Fig. 3) show that they differ significantly from each other. In AB potential the WF by coordinate $r$ inside the nucleus ($r \leq 1.5$ fm) is close to zero, the maximum value is reached at $r \approx 3.2$ fm and decreases to zero at $r \approx 8$ fm. In BFW potential inside the nucleus there is a complex oscillatory structure with two maxima and minima. The function reaches its maximum value at $r \approx 3.5$ fm and drops to zero only at $r \approx 10$ fm. Here is the state with orbital momentum $\lambda = 0$ corresponds to the 4$S$ type WF, containing two nodes at $r = 1$ and $r = 2.2$ fm. This is clearly seen in Fig. 4a, which shows the profiles at those $R$ values, when the WF reach maxima: $R = 6$ fm for AB potential and $R = 10$ fm for BFW potential. Maxima of WFs localized at $r \approx 3.2$ fm (at $R = 6$ fm for potential AB) and at $r \approx 3.5$ fm (at $R = 10$ fm for potential BFW) can be identified with the halo configuration when neutron is removed at considerable distance from the center of two α particles mass. We should note that this maximum is dominant in the WF in potential BFW.

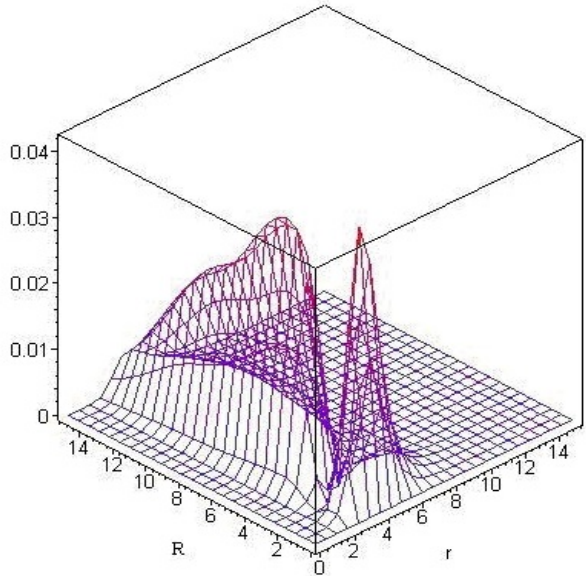 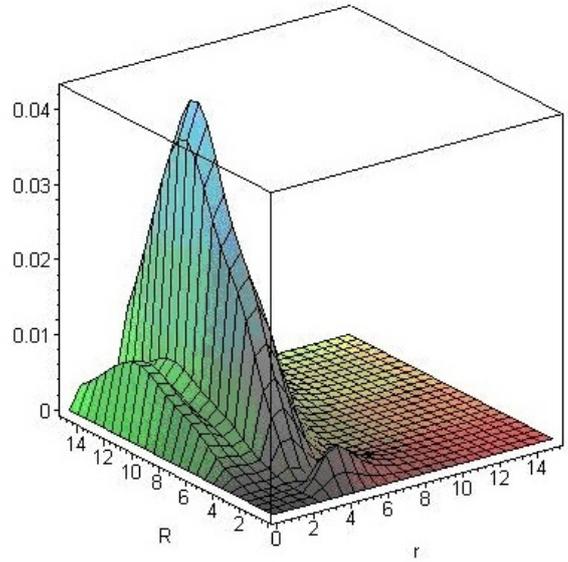

Fig. 2. Three-dimensional profile of the WF $W(r,R) = \sum_{\lambda,l,L} |\Psi^{\lambda l L}|^2 r^2 R^2$ in model 1 in the $J^\pi = 1/2^+$ state.

Fig. 3. The same as in Fig. 2 in model 2.

There is another interesting structure in both potentials by coordinate $r$ in the WF: the peak at $R \approx 1.7$ fm, $r \approx 4.5$ fm. It can be identified with the cigar-like configuration of the excited state of $^9$Be. The profile of functions for the fixed $R \approx 1.7$ fm is shown in Fig. 4b. Both peaks are comparable in absolute value in potential AB, whereas in potential BFW the first peak is less than the second by the order of magnitude.

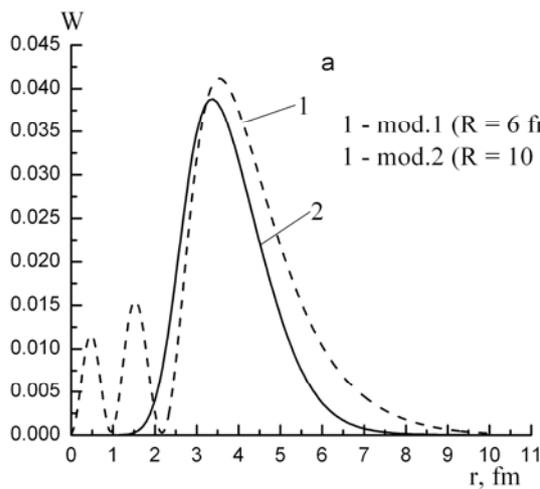 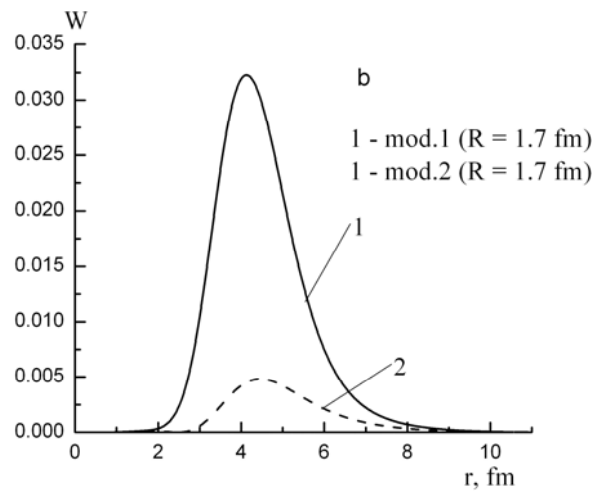

Fig. 4. Wave function $1/2^+$ profiles; a − in model 1 at $R = 6$ fm and in model 2 at $R = 10$ fm, b − in models 1 and 2 at $R = 1.7$ fm.

For large $r$ (asymptotically) in the first maximum the WF in potential AB has greater length due to the large contribution of the cigar-like configuration, in the second maximum of the WF in potential BFW (decreasing to zero at $r \approx 10$ fm).

It is also interesting to see how the WF behaves depending on $R$ − the coordinate of relative movement of neutron and the mass center of two α particles.

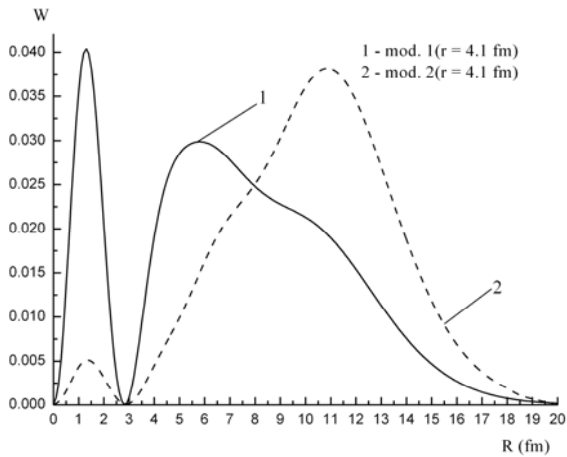
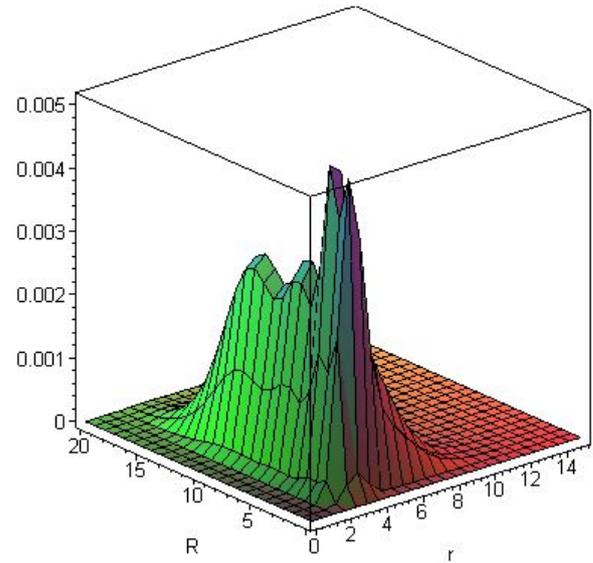

Fig. 5. Wave function $1/2^+$ profiles in models 1 and 2 at $r = 4.1$ fm.

Fig. 6. Three-dimensional profile of the WF in model 1 in $J^\pi = 3/2^+$ state.

The profiles tail at $r = 4.1$ fm (at the point, where the functions reach their maximum values) is shown in Fig. 5. In both profiles, there is a minimum at $R \approx 2.8$ fm since the WF of the $2S$ shell type with one node correspond, to the state with orbital momentum $l = 0$. Two peaks in the profiles are identified with the cigar-like configuration $R \approx 1.5$ fm, $r \approx 4.1$ fm and with triangular $R \approx 5.5$ fm, $r \approx 4.1$ fm in the potential AB and with the halo in the potential BFW $R \approx 11$ fm, $r \approx 4.1$ fm. For coordinate $R$ the more extended asymptotic behavior is observed in the WF in the potential BFW. The fact of asymptotic behavior of the WF in both potentials is greater for $R$ (than for $r$) indicates larger cohesion of 2α core in nucleus $^9$Be and much smaller cohesion of the valence nucleon.

Fig. 6 shows a three-dimensional WF profile of the excited state $J^\pi = 3/2^+$ calculated in potential AB. Having made the cutoff of the WF at several fixed $R$ values (Fig.7) we can see that for the $r$ coordinate (for all $R$ values) the WF inside the nucleus ($r \leq 1.5$ fm) is zero, it reaches the maximum value at $r \approx 3.3$–$3.5$ fm and asymptotically approaches zero at $r \approx 9$ fm.

The wave function profiles at fixed values of $r = 2.5$ and $r = 3.5$ fm demonstrate the behavior, that is similar to each other (Fig.8). If (as noted in [2, 3]) in the ground state there is a strong elongation on the coordinate $r$, in the excited states $J^\pi = 1/2^+$, $3/2^+$, on the contrary, the WFs have greatest length on the coordinate $R$. The function reaches (at $r \approx 3.5$ fm) the maximum value at $R = 3.2$ fm, it decreases very slowly, oscillating, and drops to zero only at $R \approx 18$ fm. The small first peak at $R \approx 0.5$ fm demonstrates the contribution of the cigar-like configuration, when neutron is located approximately between the two α-particles: $r = 3.5$ fm, $R \approx 0.5$ fm. However, the contribution of this configuration is small, and the triangular configuration can be realized more likely: $r = 3.5$ fm, $R \approx 3.2$ fm, or halo $r = 3.5$ fm, $R \approx 10$ fm.

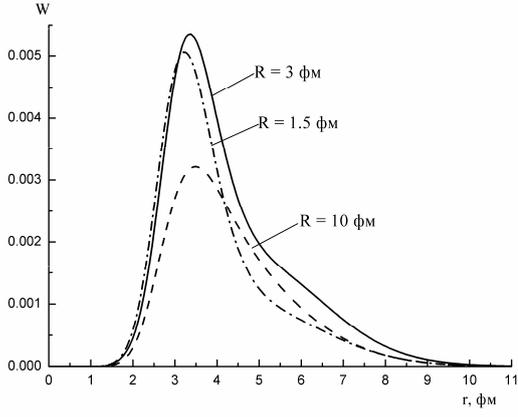 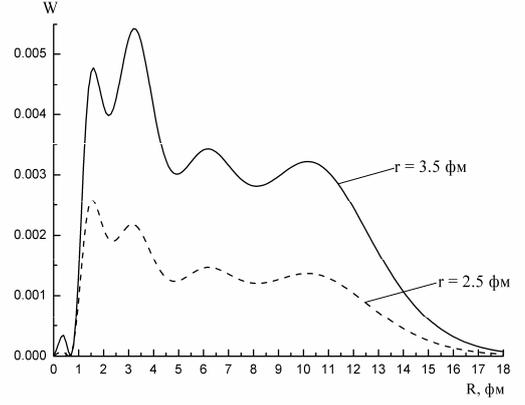

Fig. 7. Wave function $3/2^+$ profile at $R = 1.5, 3, 10$ fm.

Fig. 8. Wave function $3/2^+$ profile at $r = 2.5, 3.5$ fm.

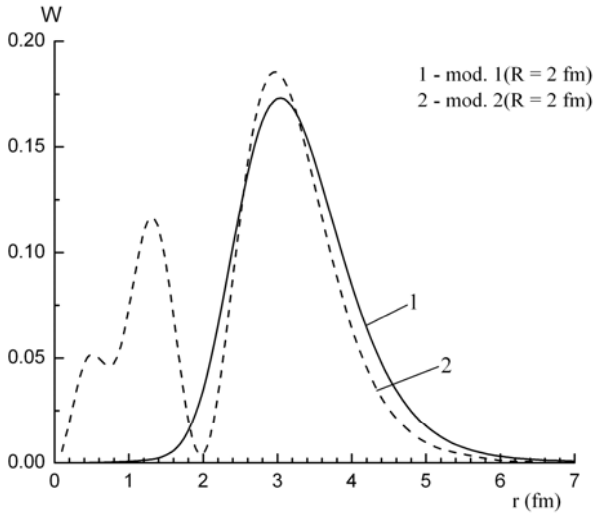

Fig. 9. Wave function profiles in the ground $J^\pi = 3/2^-$ state in models 1 and 2.

For comparison, Fig. 9 shows the function profiles of the ground state of $^9$Be (sum of all three components (3) − (5)) in the potential BFW (dashed curve) and in the potential AB (solid line), when the neutron is at $R = 2$ fm distance from the center of α-particles mass. In the potential BFW we can see the WF oscillatory structure in the inner region, the node at $r = 2$ fm, the maximum at $r = 3.2$ fm, and the rapid decrease of the function. So it is reduced to almost zero at $r = 6$ fm. In the potential AB in the inner region $r \leq 1.5$ fm the function is equal to zero, the maximum value is reached at $r = 3$ fm and it reduces asymptotically to zero at $r = 7$ fm. Comparing this behavior with the WF of the excited state (at close value of $R = 1.7$ fm in Fig. 4b) we see that the WFs in both potentials reach the maximum at $r = 4.5$ fm, and they fall down to zero only at $r = 8$–$9$ fm.

Thus, for the excited states of $^9$Be we can observe the extended neutron distribution determining its diffuse structure. This is consistent with our calculation of the rms radii: $\langle r^2 \rangle^{1/2} = 2.83$ fm for the $J^\pi = 1/2^+$ state and $\langle r^2 \rangle^{1/2} = 2.98$ fm for the $J^\pi = 3/2^+$ state.

## 3. Calculation of matrix elements for inelastic scattering

The calculation of matrix elements (amplitudes) of the elastic scattering in the Glauber theory with three-body WFs was stated in [6−8], of the inelastic scattering at the $J^\pi = 1/2^+$ level − in [28]. Discuss the main points of the matrix elements output for the inelastic scattering at the $J^\pi = 3/2^+$ level.

The matrix element of scattering in the Glauber theory is written as follows [29]:

$$M_{if}(\mathbf{q}) = \sum_{M_J M_J'} \frac{ik}{2\pi} \int d^2\boldsymbol{\rho}\, \exp(i\mathbf{q}\boldsymbol{\rho}) \delta(\mathbf{R}_A) \left\langle \Psi_f^{JM_J'} \left| \Omega \right| \Psi_i^{JM_J} \right\rangle, \qquad (7)$$

where $\boldsymbol{\rho}-$ is the impact parameter, which is a two-dimensional vector in the Glauber theory, $\mathbf{R}_A-$ the coordinate of the center of the target nucleus mass, $\Psi_i^{JM_J}, \Psi_f^{JM_J'}-$ WFs of the target nucleus in the initial and final states (formulae (3)–(6)), $\mathbf{k}, \mathbf{k}'-$ momenta of the incident and scattering proton, $\mathbf{q}-$ the transferred momentum in the reaction: $\mathbf{q} = \mathbf{k} - \mathbf{k}'$.

The Glauber operator of multiple scattering in the general form is written as a series of alternating sign one-, two-, ..., $A$-fold (where $A$ is the number of nucleons in the target nucleus) collisions of the incident proton with the nucleus nucleons [29]:

$$\Omega = 1 - \prod_{j=1}^{A}\left(1 - \omega_j(\boldsymbol{\rho} - \boldsymbol{\rho}_j)\right) = \sum_{j=1}^{A}\omega_j + \sum_{j\langle\mu}\omega_j\omega_\mu - \sum_{j\langle\mu\langle\eta}\omega_j\omega_\mu\omega_\eta + ... (-1)^{A-1}\omega_1\omega_2...\omega_A, \qquad (8)$$

where $\omega_j-$ the profile function, depending on elementary $f_{pj}(q)$-amplitude:

$$\omega_j(\boldsymbol{\rho} - \boldsymbol{\rho}_j) = \frac{1}{2\pi ik} \int d^2\mathbf{q}\, \exp\left[-i\mathbf{q}(\boldsymbol{\rho} - \boldsymbol{\rho}_j)\right] f_{pN}(q). \qquad (9)$$

Proton-nucleon amplitude is parameterized in the following standard way:

$$f_{pN} = \frac{k\sigma_{pN}}{4\pi}(i + \varepsilon_{pN})\exp\left(-\beta_{pN}q^2/2\right), \qquad (10)$$

where $\sigma_{pN}-$ the total cross section of the proton scattering on nucleon, $\varepsilon_{pN}-$ the ratio of the real part of amplitude to the imaginary one, $\beta_{pN}-$ the parameter of amplitude cone slope. The parameters of the amplitudes at different energies are given in [6, 7].

Substituting into the matrix element of the WF of $^9$Be in the $2\alpha$n-model, it is convenient to convert the operator $\Omega$ to the form adequate to this model, considering the collisions not with individual nucleons, but with α-particle clusters as structureless entities and with the remaining nucleon. In accordance with this approach, a series of multiple scattering (8) for $^9$Be can be rewritten as follows:

$$\Omega = \sum_{j=1}^{3}\omega_j - \sum_{i\langle j=1}^{3}\omega_i\omega_j + \omega_{\alpha_1}\omega_{\alpha_2}\omega_n, \qquad (11)$$

where $j = 1, 2$ enumerate $\alpha_1$ and $\alpha_2$, $j = 3$ enumerates n.

After substituting the elementary amplitude (10) in the profile function (9) and integration on $d^2\mathbf{q}$, we obtain the following:

$$\omega_j(\boldsymbol{\rho} - \boldsymbol{\rho}_j) = F_j \exp\left[-(\boldsymbol{\rho} - \boldsymbol{\rho}_j)^2 \eta_j\right], \qquad (12)$$

where

$$F_j = \frac{\sigma_{xj}}{4\pi\beta_{xj}}(i + \varepsilon_{xj}), \quad \eta_j = \frac{1}{2\beta_{xj}}. \qquad (13)$$

For further calculations it is necessary to move from single-particle coordinates of nucleons $\{\mathbf{\rho}_1, \mathbf{\rho}_2, \mathbf{\rho}_3\}$ in the operator $\Omega$ to Jacobi coordinates $\{\mathbf{r}, \mathbf{R}\}$ and the coordinate of $^9$Be mass center $-\mathbf{R}_9$. The relationship between these sets of coordinates follows from Fig.1:

$$\mathbf{r} = \mathbf{\rho}_1 - \mathbf{\rho}_2, \quad \mathbf{R} = \frac{\mathbf{\rho}_1 + \mathbf{\rho}_2}{2} - \mathbf{\rho}_3, \quad \mathbf{R}_9 = \frac{1}{9}(4\mathbf{\rho}_1 + 4\mathbf{\rho}_2 + \mathbf{\rho}_3). \tag{14}$$

Inverse transformations will give the following:

$$\mathbf{\rho}_1 = \mathbf{R}_9 - \frac{\mathbf{R}}{9} + \frac{\mathbf{r}}{2}; \quad \mathbf{\rho}_2 = \mathbf{R}_9 + \frac{\mathbf{R}}{9} - \frac{\mathbf{r}}{2}; \quad \mathbf{\rho}_3 = \mathbf{R}_9 - \frac{8}{9}\mathbf{R}. \tag{15}$$

As it was shown in [30], after some transformations, the operator $\Omega$ in Jacobi coordinates can be written as

$$\Omega = (\mathbf{G}\ \mathbf{H}) = \sum_{k=1}^{7} G_k H_k, \tag{16}$$

where the summation by index $k$ means the summation by the scattering multiplicity: $k = 1–3$ are one-time collisions, $k = 4–6$ are two-times collisions, $k = 7$ is three-times collision. Here the $\mathbf{G}$ is 7-dimensional vector with components

$$\mathbf{G} = (G_1, G_2, ..., G_7) = (F_\alpha, F_\alpha, F_n, -F_\alpha F_\alpha, -F_\alpha F_n, -F_\alpha F_n, F_\alpha F_\alpha F_n). \tag{17}$$

The components of vector $\mathbf{H} = (H_1, H_2, ... H_7)$ are expressed through the exponential function from the elements $a_k, b_k, ...$ according to the formula:

$$H_k = \exp(-a_k \mathbf{\rho}_\perp^2 - b_k \mathbf{R}_\perp^2 - c_k \mathbf{r}_\perp^2 + d_k \mathbf{\rho}_\perp \mathbf{R}_\perp + l_k \mathbf{\rho}_\perp \mathbf{r}_\perp + f_k \mathbf{R}_\perp \mathbf{r}_\perp), \tag{18}$$

where

$$a_k = (\eta_\alpha, \eta_\alpha, \eta_n, 2\eta_\alpha, (\eta_\alpha + \eta_n), (\eta_\alpha + \eta_n), (2\eta_\alpha + \eta_n)),$$

$$b_k = \frac{1}{81}(\eta_\alpha, \eta_\alpha, 64\eta_n, 2\eta_\alpha, (\eta_\alpha + 64\eta_n), (\eta_\alpha + 64\eta_n), (2\eta_\alpha + 64\eta_n)),$$

$$c_k = \frac{1}{2}\left(\frac{\eta_\alpha}{2}, \frac{\eta_\alpha}{2}, 0, \eta_\alpha, \frac{\eta_\alpha}{2}, \frac{\eta_\alpha}{2}, \eta_\alpha\right),$$

$$d_m = \frac{2}{9}(\eta_\alpha, \eta_\alpha, 8\eta_n, 2\eta_\alpha, (2\eta_\alpha + 8\eta_n), (2\eta_\alpha + 8\eta_n), (2\eta_\alpha + 8\eta_n)),$$

$$l_m = (-\eta_\alpha, \eta_\alpha, 0, 0, -\eta_\alpha, \eta_\alpha, 0), \quad f_m = \frac{1}{9}(\eta_\alpha, -\eta_\alpha, 0, 0, \eta_\alpha, -\eta_\alpha, 0).$$

By substituting the WF (3)−(6) in the formula (7), we will obtain the following nonzero matrix elements:

$$M_{if}^{(\lambda lL)}(\mathbf{q}) = M_{if}^{(011)}(\mathbf{q}) + M_{if}^{(211)}(\mathbf{q}) + M_{if}^{(212)}(\mathbf{q}), \tag{19}$$

where

$$M_{if}^{(011)}(\mathbf{q}) = \frac{ik}{2\pi} \int \exp(i\mathbf{q}\boldsymbol{\rho})\delta(\mathbf{R}_9)d\boldsymbol{\rho}\langle\Psi_{022}^{\lambda lL}|\Omega|\Psi_{011}^{\lambda lL}\rangle = \frac{ik}{2\pi} \times$$

$$\times \int \exp(i\mathbf{q}\boldsymbol{\rho})\delta(\mathbf{R}_9)d\boldsymbol{\rho}\left\langle \frac{1}{\sqrt{4\pi}}\chi_{\frac{1}{2}M_S'}\sum_{ij}\left\langle 2M_L'\frac{1}{2}M_S'\bigg|\frac{3}{2}M_J'\right\rangle Y_{2M_L'}(\Omega_R)R^2\sum_{i'j'}C_{i'j'}^{02}\exp(-\alpha_i'r^2 - \beta_j'R^2)\right.$$

$$\left.\bigg|\Omega\bigg|\sum\frac{1}{\sqrt{4\pi}}\left\langle 1M_L\frac{1}{2}M_S\bigg|\frac{3}{2}M_J\right\rangle Y_{1m}(\Omega_R)\chi_{\frac{1}{2}M_S}R\sum_{\nu\mu}C_{ij}^{01}\exp(-\alpha_\nu r^2 - \beta_\mu R^2)\right\rangle$$

(20)

$$M_{if}^{(211)}(\mathbf{q}) = \frac{ik}{2\pi} \int \exp(i\mathbf{q}\boldsymbol{\rho})\delta(\mathbf{R}_9)d\boldsymbol{\rho}\langle\Psi_{022}^{\lambda lL}|\Omega|\Psi_{211}^{\lambda lL}\rangle = \frac{ik}{2\pi} \times$$

$$\times \int \exp(i\mathbf{q}\boldsymbol{\rho})\delta(\mathbf{R}_9)d\boldsymbol{\rho}\left\langle \frac{1}{\sqrt{4\pi}}\chi_{\frac{1}{2}M_S'}\sum_{ij}\left\langle 2M_L'\frac{1}{2}M_S'\bigg|\frac{3}{2}M_J'\right\rangle Y_{2M_L'}(\Omega_R)R^2\sum_{i'j'}C_{i'j'}^{02}\exp(-\alpha_i'r^2 - \beta_j'R^2)\right.$$

$$\left.\bigg|\Omega\bigg|\sum_{M_LM_S\mu}\left\langle 1M_L\frac{1}{2}M_S\bigg|\frac{3}{2}M_J\right\rangle\langle 2\mu 1m|1M_L\rangle Y_{2\mu}(\Omega_r)Y_{1m}(\Omega_R)\chi_{\frac{1}{2}M_S}\sum_{\nu\varepsilon}C_{\nu\varepsilon}^{21}R\cdot r^2\exp(-\alpha_\nu r^2 - \beta_\varepsilon R^2)\right\rangle$$

(21)

$$M_{if}^{(212)}(\mathbf{q}) = \frac{ik}{2\pi} \int \exp(i\mathbf{q}\boldsymbol{\rho})\delta(\mathbf{R}_9)d\boldsymbol{\rho}\langle\Psi_{022}^{\lambda lL}|\Omega|\Psi_{212}^{\lambda lL}\rangle = \frac{ik}{2\pi} \times$$

$$\times \int \exp(i\mathbf{q}\boldsymbol{\rho})\delta(\mathbf{R}_9)d\boldsymbol{\rho}\left\langle \frac{1}{\sqrt{4\pi}}\chi_{\frac{1}{2}M_S'}\sum_{ij}\left\langle 2M_L'\frac{1}{2}M_S'\bigg|\frac{3}{2}M_J'\right\rangle Y_{2M_L'}(\Omega_R)R^2\sum_{i'j'}C_{i'j'}^{02}\exp(-\alpha_i'r^2 - \beta_j'R^2)\right.$$

$$\left.\bigg|\Omega\bigg|\sum_{M_LM_S\mu}\left\langle 2M_L\frac{1}{2}M_S\bigg|\frac{3}{2}M_J\right\rangle\langle 2\mu 1m|2M_L\rangle Y_{2\mu}(\Omega_r)Y_{1m}(\Omega_R)\chi_{\frac{1}{2}M_S}\sum_{\gamma\zeta}C_{\gamma\zeta}^{21}R\cdot r^2\exp(-\alpha_\gamma r^2 - \beta_\zeta R^2)\right\rangle$$

(22)

It should be noted, that all vectors in the operator $\Omega$ lying in the xy plane are two-dimentional. The vectors with the same name in the WF are three-dimensional, so we perform the integration of the matrix element (1) in the Cartesian coordinate system. In the WF (18)−(20) let's move from regular sectorial harmonic to polynomials according to the formula from [31]:

$$\mathbf{R}^l Y_{lm}(\Omega_R) = \sqrt{\frac{2l+1}{4\pi}(l+m)!(l-m)!}\sum_{u,v,w}\frac{1}{u!v!w!}\left(-\frac{R_x+iR_y}{2}\right)^u\left(\frac{R_x-iR_y}{2}\right)^v R_z^w, \qquad (23)$$

where u,v,w are positive integral numbers: u+v+w = l, u−v=m; $R_x$, $R_y$, $R_z$ are projections of vector $\mathbf{R}$ on the axis of the Cartesian coordinate system.

Taking the formula (23) in note, we calculate the sum in the matrix elements (20)–(22).

$$\sum_{M_LM_S}\left\langle 2M_L'\frac{1}{2}M_S'\bigg|\frac{3}{2}M_J'\right\rangle\left\langle 1M_L\frac{1}{2}M_S\bigg|\frac{3}{2}M_J\right\rangle R^2 Y_{2M_L'}(\Omega_R)R\cdot Y_{1m}(\Omega_R) = \frac{3\sqrt{5}}{2\pi}R_x R_y^2, \quad (24)$$

$$\sum_{M_LM_SM_L'M_S'\mu m}\left\langle 2M_L'\frac{1}{2}M_S'\bigg|\frac{3}{2}M_J'\right\rangle\left\langle 1M_L\frac{1}{2}M_S\bigg|\frac{3}{2}M_J\right\rangle\langle 2\mu 1m|1M_L\rangle R^2 Y_{2M_L'}(\Omega_R)R\cdot Y_{1m}(\Omega_R)r^2 Y_{2\mu}(\Omega_r) =$$

$$= \frac{ik\sqrt{3}(\sqrt{3}+1)}{(2\pi)^3\sqrt{2}}\{(r_z^2 + r_x - 2r_y)R_y - 3r_x r_y R_x\}, \qquad (25)$$

$$\sum_{M_L M_S M'_L M'_S \mu m} \left\langle 2M'_L \frac{1}{2} M'_S \bigg| \frac{3}{2} M'_J \right\rangle \left\langle 2M_L \frac{1}{2} M_S \bigg| \frac{3}{2} M_J \right\rangle \langle 2\mu 1m | 2M_L \rangle R^2 Y_{2M'_L}(\Omega_R) R \cdot Y_{1m}(\Omega_R) r^2 Y_{2\mu}(\Omega_r) =$$

$$= \frac{ik3\sqrt{2}(\sqrt{3}-1)}{(2\pi)^3} R_x R_y \left( r_z^2 R_x - r_y^2 R_x + r_x r_y R_y \right). \qquad (26)$$

By substituting the operator (16) and the calculated sums (24)–(26) in the matrix elements (20)–(22), we see that the integrals over the projections $\rho_x$, $\rho_y$, $R_x$, $R_y$, $R_z$, $r_x$, $r_y$, $r_z$ have the form of Gaussian integrals (or Euler-Poisson integrals) and they can be calculated analytically. It is important to note that the use of this approach, when the WFs are recorded as an expansion in gaussoids and the operator is represented in the form conjugated the WF. It becomes possible to consider all the multiplicity of scattering on clusters and nucleons of the nucleus and to calculate the matrix elements without any simplifications, and therefore without loss of precision.

The differential cross section is a square of the matrix element module:

$$\frac{d\sigma}{d\Omega} = \frac{1}{2J+1} \sum_{M_J M'_J} |M_{if}(\mathbf{q})|^2. \qquad (27)$$

## 4. Differential cross section of the inelastic scattering

Figs. 10, 11 show the calculation of the differential cross section of the inelastic p$^9$Be scattering with different model WFs of $^9$Be. The solid and dashed curves indicate the calculation with WFs in models 1, 2, the dash-dot line indicates the calculation with oscillator WF, dotted line indicates the calculation from [20]. The differential cross section at the levels $J^\pi = 1/2^+$, $3/2^+$, which we use to compare our calculation, is measured in the experiment made in the Cyclotron Laboratory at Indiana University [20] at $E_p = 180$ MeV.

In Fig. 10 (scattering at the level $J^\pi = 1/2^+$) we can see that the differential cross section with three-body wave function in models 1 and 2, although similar to each other, but different somewhat at mean and large scattering angles (momentum transfers). The minimum in the differential cross section at $\theta \to 0°$ is caused by the orthogonality of the WF of the initial and final states of $^9$Be. Furthermore, the cross sections rapidly increase to a maximum and then decrease monotonically, with increase of scattering angle. The contribution to the differential cross section at small angles depends on the behavior of the WF asymptotically. The abovementioned analysis of WF profiles shows that in the excited states with different values of the coordinates $R$ and $r$, the lengths of the WFs in models 1 and 2 are different. According to the coordinate r at the first maximum (due to large contribution of the cigar-like configuration) more extended tail is observed in the WF of model 1 (Fig. 4b), in the second maximum − on the contrary, in the WF of model 2 (Fig. 4a). According to coordinate $R$, the extended tail of the WF is observed in model 2 (Fig. 5). In the ground state the longer function is in model 1 (Fig. 9). It is not possible to make the unambiguous conclusion about the greater length of the WF in any model.

Discrepancies in the differential cross section at medium (and large) scattering angles point to different behavior of the WF in the inner region of nucleus. The wave function in model 1 in the center of the nucleus has a repusive core (Fig. 4a, b). Its contribution to the cross section at moderate angles ($\theta > 35°$) is less and the cross section with the WF decreases in model 1 (solid curve) faster than the WF in model 2 (dashed curve).

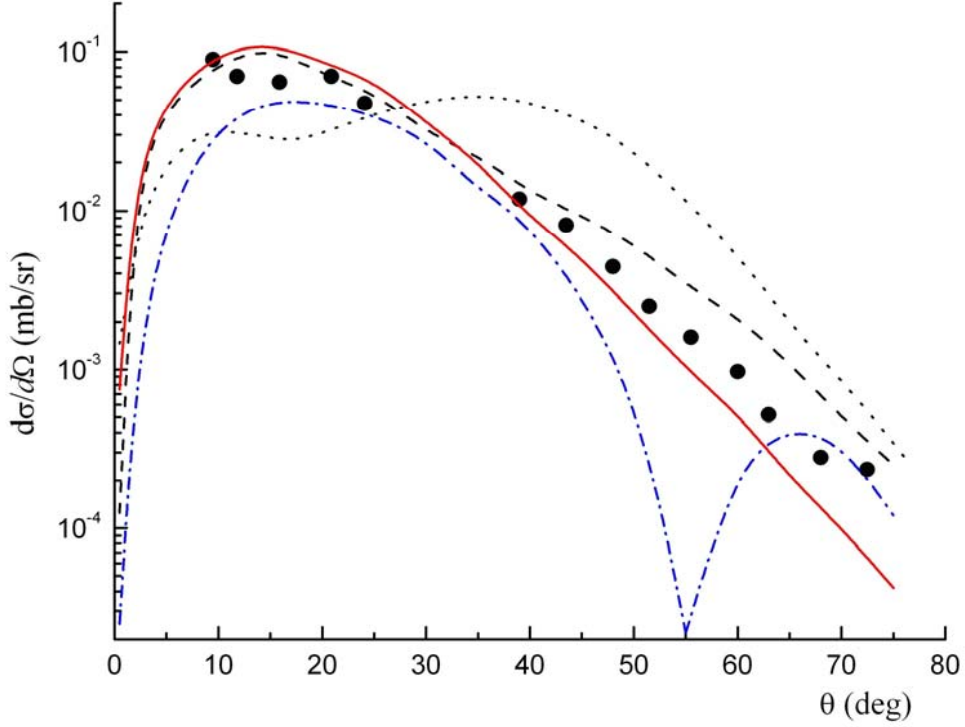

Fig. 10. The differential cross section for the inelastic p$^9$Be scattering to the level $J^\pi = 1/2^+$ at $E$ = 180 MeV. The solid, dashed and dashed-dot lines are our calculation with the WF in models 1, 2 and with a shell model. The experimental points and the dotted line are from [20].

Also, we have done the calculation with the WF in the shell model, where the WF of the ground state is $\Psi_i = 0.899\{41\}^{22}P - 0.387\{41\}^{22}D$ [25], and the WF of the excited state is $\Psi_f = {}^{22}S$. If we consider only one dominant component of the WF, this is the differential cross section (dash-dotted curve) at small angles tends to zero faster (than with three-particles functions), its absolute value in the maximum of the cross section is less than the experimental one and the minimum is observed at $\theta \sim 55°$ in the cross-section. This behavior of the cross section is stipulated by the fact that the oscillatory WF asymptotically decreases faster than the three-body one (this is reflected in the behavior of the cross section at low $\theta$), and it has the node (providing minimum in the cross section). However, if we would also take into account in the WF of the ground and excited states the $D$ component then its contribution (e.g., as shown by calculation for $^6$Li [6]) would fill the minimum and smooth the cross-section.

A similar pattern of scattering to the level $J^\pi = 3/2^+$ is presented in Fig. 11. The wave function, calculated in model 1, has the extended asymptotics by coordinate $r$ (extending up to ~9 fm, Fig. 7) and even more in the coordinate $R$ (extending up to ~18 fm, Fig. 8). It leads to a rapid increase of the cross section at small angles. The maximum of the calculated differential cross section is close to the maximum of the experimental one. However at $\theta > 40°$ it falls off more rapidly, than the experimental one. The cross section with the shell model WF $\Psi_f = {}^{22}D$ correlates worse with the experimental data throughout the angular range, although there is no minimum in the cross section, as in the previous case, because there is no node in the WF.

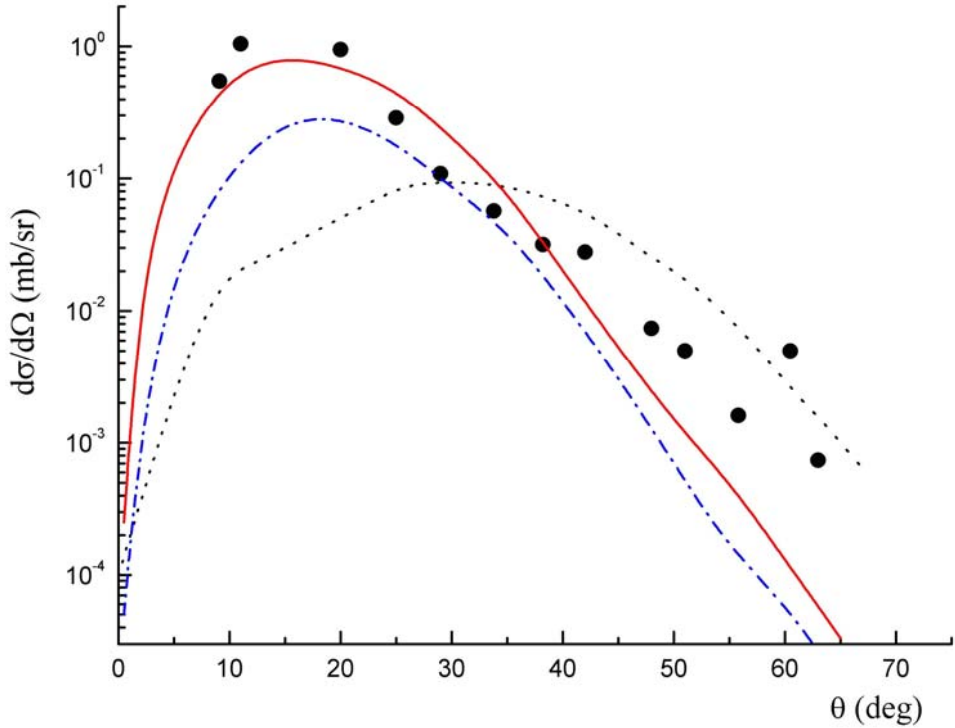

Fig. 11. The same as in Fig. 10, but only to the level $J^\pi = 3/2^+$.

It should be noted that the Glauber theory has significant limitations on the energy and angular range of the scattered particles. Since the energy of the incident particles is not too large, the results are only valid for the front scattering angles. Calculation at large angles is beyond the accuracy of the Glauber theory.

For the purpose of comparison we present in two figures with the results of the differential cross section calculation (dotted curves) in the DWBA with an effective interaction, which depends on density and based on the Paris potential with the WF in the shell model [20]. However, the differential cross sections are much smaller for small angles (momentum transfers) and significantly large for larger angles. The authors explain this that "the oscillator WFs in the spherical basis inadequately describe the $1\hbar\omega$ excitations of this strongly deformed nucleus" [21].

It is interesting to note, that the differential cross sections to the levels of negative parity $J^\pi = 1/2^-, 3/2^-, 5/2^-$ in [20] are calculated very accurately, and exactly agree with the experiment, where the differential cross sections to similar levels of the positive parity, in contrast, they are very different from the experiment.

Probably, this is reflection of halo-structure of levels with positive parity. The predictions about excited halo-states of nuclei were expressed in [32]. In the experiment on $^{13}C$ [33], the halo-state in the first excited state $J^\pi = 1/2^+$, $E^* = 3.089$ MeV was observed. In [34, 35] a modified diffraction model for determining the nuclear radii for such short-lived excited states was proposed and applied it to determine the diffraction radii of $^9Be$ and $^{11}Be$ in the excited states [9, 10]. Meanwhile significantly enlarged radii (approximately for 1.2 fm [10]) were found for the $J^\pi = 1/2^+$ ($E^* = 1.68$ MeV) and $J^\pi = 5/2^+$, ($E^* = 3.05$ MeV) states of $^9Be$, that confirms our conclusion.

## 5. Conclusion

The analysis of the WF of $^9Be$ in the $2\alpha n$ model showed that in the excited states $J^\pi = 1/2^+, 3/2^+$ the nucleus has a more extended, diffuse structure in both models compared with the WF of the ground state. Comparison of WFs in models 1 and 2 (for the $J^{\pi} = 1/2^+$ state) shows significantly different behavior in the inner region (close to zero in model 1 and oscillating in model 2) and about the same, slowly decaying asymptotically.

Having written the operator $\Omega$ in the form conjugated with the 2αn model WF, we were able to calculate the matrix elements of the inelastic scattering analytically, taking into account all the multiplicity of scattering on α clusters and the nucleon.

The cross-sections calculated in the framework of the Glauber theory of multiple scattering are correctly consistent with the available experimental data in the front angles. The analysis of WF profiles made it possible to connect them with the behavior of the cross section and to identify the effect of the contribution of different WF regions on the differential cross section. From all the calculated curves the solid curve is closer to the experimental one, calculated with the wave functions in the model 1. Thus, the dynamic characteristic, the differential cross section, calculated by us, confirms the findings of the authors [2,3] that the most adequate description of all characteristics is obtained with WF calculated in a small $\lambda$-dependent potential AB with a repulsive core, which suppresses the oscillations of the wave function in the inner region of the nucleus. We also obtained the similar result for elastic p$^9$Be scattering [8].

## Acknowledgments

The authors express their gratitude to Voronchev V.T. for providing the $^9$Be wave functions in the excited states and to the Committee of Science of the Ministry of Education and Science of the Republic of Kazakhstan for financial support (grant 1124/GF).